\documentclass{iaus}
\usepackage{graphicx}

\title[RGB Stars in I Zw 18] 
{A New Deep HST/ACS CMD of I~Zw~18: Evidence for Red 
Giant Branch Stars}

\author[Aloisi et al.]   
{A.~Aloisi$^{1,2}$, F.~Annibali$^1$, J.~Mack$^1$, M.~Tosi$^3$, 
R.~van der Marel$^1$, G.~Clementini$^3$, R.~A.~Contreras$^3$, 
G.~Fiorentino$^3$, M.~Marconi$^4$, I.~Musella$^4$, A.~Saha$^5$}

\affiliation{$^1$Space Telescope Science Institute, Baltimore, USA 
\break email: aloisi@stsci.edu\\[\affilskip]
$^2$ On Assignment from the Space Telescope Division of the European 
Space Agency\\[\affilskip]
$^3$INAF - Osservatorio Astronomico di Bologna, Bologna, Italy\\[\affilskip]
$^4$INAF - Osservatorio Astronomico di Capodimonte, Napoli, Italy\\[\affilskip]
$^5$National Optical Astronomy Observatories, Tucson, USA\\[\affilskip]
}

\pubyear{2007}
\volume{241}
\pagerange{}
\date{}
\jname{Stellar Populations as Building Blocks of Galaxies}
\editors{R.~F.~ Peletier, \& A.~Vazdekis, eds.}
\begin{document}

\maketitle

\begin{abstract}
We present results from new deep HST/ACS photometry of I~Zw~18, the
most metal-poor blue compact dwarf galaxy in the nearby universe. It
has been previously argued that this is a very young system that
started forming stars only $\lesssim 500$ Myr ago, but other work has
hinted that older ($\gtrsim 1$ Gyr) red giant branch (RGB) stars may
exist in this galaxy. Our deeper data indeed reveal evidence for an
RGB.  Underlying old ($\gtrsim 1$ Gyr) populations are therefore
present in even the most metal-poor systems, implying that star
formation started at $z \gtrsim 0.1$. The RGB tip (TRGB) magnitude and
the properties of Cepheid variables identified from our program
indicate that I~Zw~18 is farther away ($D = 19.0 \pm 1.8$ Mpc) than
previously believed.

\keywords{galaxies: dwarf, irregular, starburst, stellar content,
distances and redshifts, Cepheids}
\end{abstract}

\firstsection 

\section{Introduction}
Within the framework of hierarchical formation, dwarf ($M$ $\lesssim$
10$^9$ M$_{\odot}$) galaxies are the first systems to collapse and
start forming stars, supplying the building blocks for the formation
of more massive galaxies through merging and accretion. As remnants of
this process, present-day dwarfs may have been sites of the earliest
star-formation (SF) activity in the universe.  However, the most
metal-poor ($12 + \log (O/H) \lesssim 7.6$, corresponding to $Z$
$\lesssim$ 1/20 Z$_{\odot}$) dwarf irregular (dIrr) and blue compact
dwarf (BCD) galaxies have been repeatedly pointed out as good
candidate ``primeval'' galaxies in the nearby universe, with possible
ages less than 100-500 Myr (e.g., Izotov \& Thuan 1999). If some of
these objects turn out to be young galaxies, their existence would
support the view that SF in low-mass systems has been inhibited until
the present epoch (e.g., Babul \& Rees 1992). The discovery of a
nearby primordial galaxy would thus have revolutionary cosmological
implications. On the other hand, the lack of such systems would
provide strong constraints on the chemical and physical evolution of
low-metallicity dIrr/BCDs.

The only direct way to unambiguously infer the evolutionary status of
a metal-poor dIrr/BCD galaxy is to resolve it into stars with deep HST 
observations, and study stellar features in the color-magnitude diagram 
(CMD). The brightest of these features that contains stars of significant 
age is the Red Giant Branch (RGB), formed by low-mass stars with ages 
$\sim 1$-13 Gyr that are burning H in a shell around a He core. In the 
last 15 years all metal-poor dIrr galaxies in the Local Group and BCDs 
with $D \lesssim 15$ Mpc have been imaged with HST. Interestingly, an 
RGB has been detected in all those galaxies for which photometric data 
exist that go deep enough to reach the RGB tip (TRGB, brightest phase 
of the RGB) in the CMD (e.g., SBS~1415+437, Aloisi et al.~2005, and 
references therein).\looseness=-2

The only possible exception so far is the BCD
I~Zw~18. With an oxygen abundance $12 + \log (O/H) = 7.2$, 
corresponding to $Z = 1/50$ Z$_{\odot}$, I~Zw~18 is the ``prototype'' 
member of its class, and the most metal-poor galaxy in the nearby 
universe. Many HST studies have focused on the evolutionary status 
of I~Zw~18. After other groups had already resolved the brightest 
individual stars in the galaxy, our group was the first to go deep 
enough to detect asymptotic giant branch (AGB) stars in HST/WFPC2 
images with ages of at least several hundreds Myr and possibly up 
to a few Gyrs (Aloisi et al.~1999). These results were confirmed by 
\"Ostlin (2000) through deep HST/NICMOS imaging. More recently, 
Izotov \& Thuan (2004) presented new deep HST/ACS observations. 
Their $I$ vs.~$V-I$ CMD shows no sign of an RGB, from which they 
concluded that the most evolved (AGB) stars are not older than 500 
Myr and that I~Zw~18 is a bona fide young galaxy. This result was 
subsequently challenged by Momany et al.~(2005) and our group (Tosi 
et al.~2006) based on a better photometric analysis of the same data. 
This showed that many red sources do exist at the expected position 
of an RGB, and that their density in the CMD drops exactly where a 
TRGB would be expected. However, the small number statistics, the 
large photometric errors, and the incompleteness at the TRGB, did 
not allow a conclusive statement about the possible existence of 
an RGB in I~Zw~18.
\looseness=-2

\section{New HST/ACS Observations}
We were awarded 24 additional orbits with ACS over a three-month
period starting in October 2005 (GO program 10586, PI Aloisi) to
better understand the evolutionary status of I~Zw~18. The observations
were obtained in 12 different epochs in F606W and F814W to: (1) build
a deeper CMD to search for RGB stars; (2) detect and characterize
Cepheids at the lowest metallicity available in the local universe;
and (3) use both the Cepheids and a possible TRGB detection to
determine an accurate distance to I~Zw~18. In this paper we will
present the preliminary results from the new deep
photometry. Preliminary results from the study of the variable stars
in this metal-poor galaxy are discussed in Fiorentino et al.~(2007, in
this volume). Detailed descriptions of the observational program, the
implications for the evolutionary state and distance of I~Zw~18, and
the results on Cepheid variability will be presented in forthcoming
papers in the refereed literature.\looseness=-2

PSF-fitting photometry was performed on deep images that were obtained
by combining the exposures in each filter with MultiDrizzle. After
application of CTE and aperture corrections, the count rates were
transformed to Johnson-Cousins $V$ and $I$ magnitudes. Values shown
and discussed hereafter are corrected for $E(B-V) = 0.032$ mag of 
Galactic foreground extinction, but not for any extinction intrinsic 
to I~Zw~18. The archival ACS data in F555W and F814W 
(GO program 9400, PI Thuan) were also
re-processed in a similar manner. The two ACS datasets were then
combined using several different approaches and rejection schemes.
The results discussed here were obtained by demanding that stars
should be detected in all the four deep images ($V$ and $I$ for both
datasets). At the expense of some depth, this approach has the
advantage of minimizing the number of false detections and therefore
providing relatively ``clean'' CMDs.

\begin{figure}[t]
\begin{center}
\vspace{-0.1truecm}\includegraphics[width=0.93\hsize]{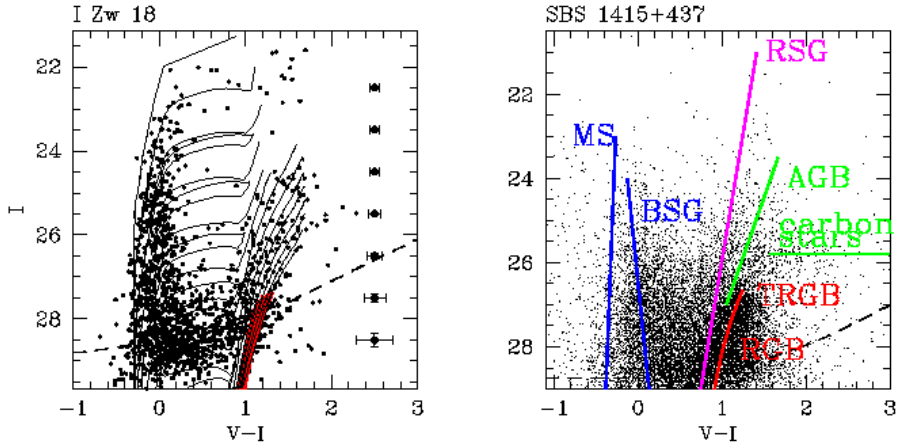}
\caption{(\textit{a}) HST/ACS CMD for I~Zw~18. Median photometric
errors at $V-I = 1$ (determined by comparison of measurements from
GO-9400 and GO-10586) are shown as function of $I$-band magnitude on
the right side of the panel. Padua isochrones from 5.5 Myr to 10 Gyr
are overlaid, with the RGB phase for isochrones from 1.7 to 10 Gyr
colored red. The isochrones have metallicity $Z = 0.0004$ (as inferred
from the HII regions of I~Zw~18) and are shown for the distance $D=19$
Mpc ($m-M = 31.39$). The CMD shown here includes stars in both the
main and secondary bodies of I~Zw~18. (\textit{b}) HST/ACS CMD for
SBS~1415+437 (Aloisi et al.~2005). The main evolutionary sequences
seen in the data are indicated in approximate sense as colored
straight lines: main sequence (MS), blue supergiants (BSG), red
supergiants (RSG), the red giant branch (RGB) with its tip (TRGB), the
asymptotic giant branch (AGB), and carbon stars. Both CMDs are
corrected for Galactic foreground extinction. Dashed lines are
estimates of the 50\% completeness level. The vertical axes of the
panels are offset from each other by 0.66 mag, which is the difference
in distance modulus between the galaxies (see Fig.~2). Some $\sim 10$
times more stars were detected in SBS 1415+437, owing to its smaller
distance; stars for this galaxy are shown with smaller symbols. The
faint red stars in both galaxies indicate that these metal-poor BCD
galaxies started forming stars $\gtrsim 1$ Gyr ago.}
\end{center}
\end{figure}

\begin{figure}[t]
\begin{center}
\includegraphics[width=0.80\hsize]{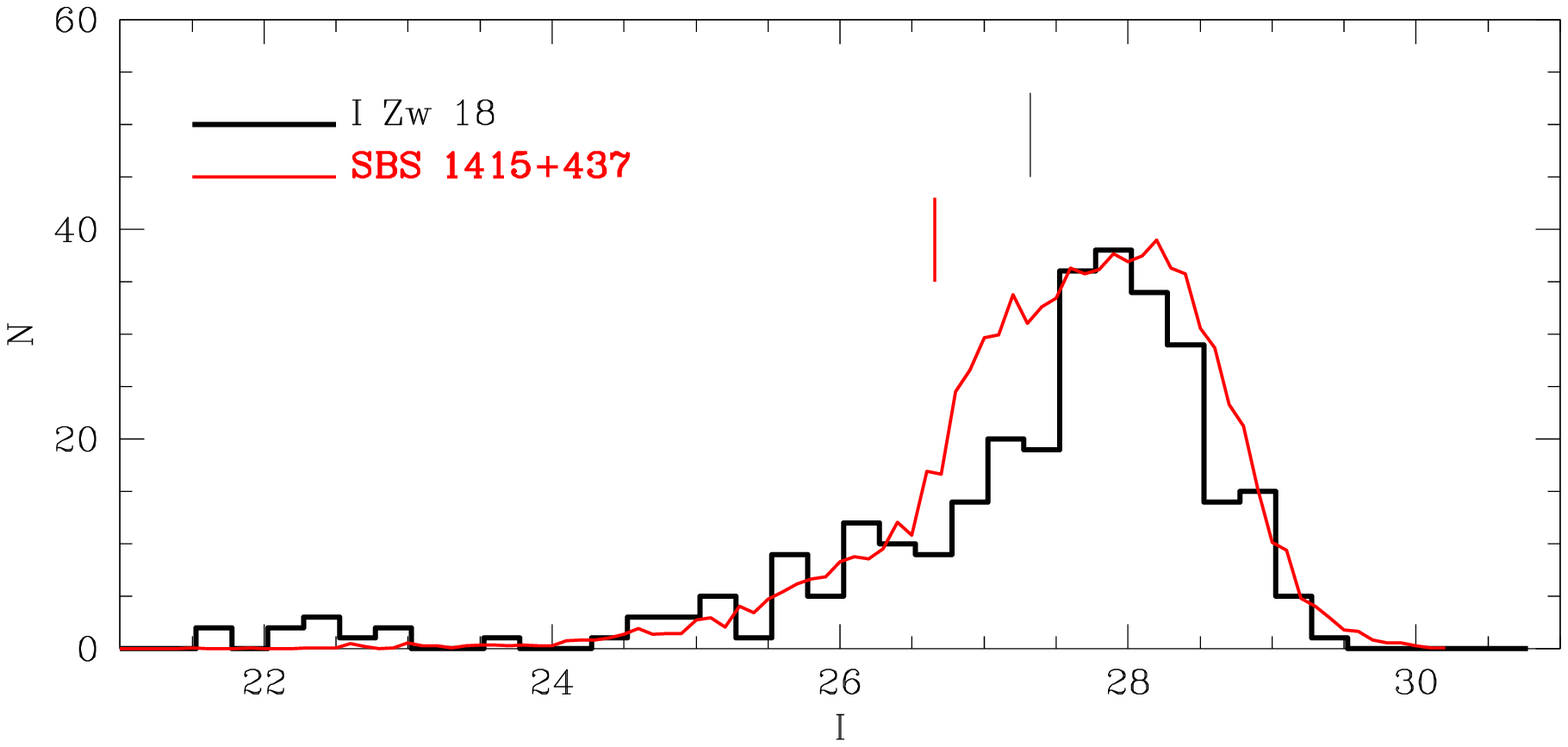}
\caption{$I$-band LFs for stars with red colors in the range $V-I =
0.75$--1.5 mag, inferred from the CMDs in Fig.~1. Normalizations are
arbitrary. Vertical marks indicate the positions of the TRGB, as
determined using a Savitzky-Golay filtering technique developed by one
of us (see Cioni et al.~2000). At these magnitudes there is a steep LF
drop towards brighter magnitudes, due to the end of the RGB sequence. By
contrast, the LF drop towards fainter magnitudes at $I \gtrsim 28$ mag
is due to incompleteness in both cases. Apart from a shift $\Delta
(m-M) \approx 0.66$, these metal-poor BCD galaxies have very similar
LFs.}
\end{center}
\end{figure}

\section{Results and Interpretation}
Fig.~1a shows the resulting $I$ vs.~$V-I$ CMD of I~Zw~18. The CMD
shows faint red stars exactly at the position where an RGB would be
expected (see the Padua isochrones overplotted in the
figure). Figure~2 shows the luminosity function (LF) of the red
stars. It shows a sharp drop towards brighter magnitudes, exactly as
would be expected from a TRGB. The magnitude of the discontinuity, $I
= 27.32 \pm 0.10$, implies a distance modulus $m-M = 31.39 \pm 0.20$
(e.g., Bellazzini et al.~2001), i.e., $D = 19.0 \pm 1.8$ Mpc. This 
assumes that the evolved RGB stars have negligible intrinsic 
extinction. The TRGB distance is
consistent with the preliminary analysis of the first few Cepheid
variables identified by our program. Comparison of their Wesenheit
relation to the Wesenheit relation inferred from new theoretical
Cepheid pulsation models at the metallicity of I~Zw~18 yields $m-M =
31.28 \pm 0.26$ (Fiorentino et al.~2007, in this volume). This
agreement further supports our interpretation of the LF drop in
Figure~2 as a TRGB feature.\looseness=-2

The evidence for an RGB in I~Zw~18 is further strengthened by
comparison to another BCD, SBS~1415+437, observed by us with a similar
HST/ACS set-up (Aloisi et al.~2005). This galaxy is not quite as metal
poor as I~Zw~18 ($12 + \log (O/H) = 7.6$) and is somewhat nearer at $D
\approx 13.6$ Mpc. But taking into account the differences in distance
and completeness, the CMDs of these galaxies look very similar. Since
SBS 1415+437 has an unmistakable RGB sequence, this suggests that such
an RGB sequence exists in I~Zw~18 as well. It has previously been
suggested that I~Zw~18 has no stars with ages $\gtrsim 500$ Myr. It is
therefore interesting to ask the question whether there exist Star
Formation Histories (SFHs) which do not have stars older than $\sim 1$
Gyr (and by extension, do not have an RGB), yet which can still fit:
(1) the observed number of faint red stars in I~Zw~18 ($V-I \gtrsim
1.0$, $I \gtrsim 27.3$); and (2) the observed LF drop-off at $I
\approx 27.3$. Preliminary results of SFH modeling that we have
performed indicate that such SFHs do not exist.

\section{Conclusions}
We have obtained new deep HST/ACS observations of I~Zw~18 that provide
improved insight into the evolutionary status of this benchmark
metal-poor BCD. Our results indicate that this galaxy contains RGB
stars, in agreement with findings for other local metal-poor BCDs
studied with HST. Underlying old ($\gtrsim 1$ Gyr) populations are
therefore present in even the most metal-poor systems, so they must
have started forming stars at $z \gtrsim 0.1$. Deeper studies (well
below the TRGB) will be needed to pinpoint the exact onset of star
formation in these galaxies. We find that at $D = 19.0 \pm 1.8$ Mpc,
I~Zw~18 is more distant than the values $\sim 15$ Mpc that have often
been assumed in previous work. This may explain why it has remained
difficult for so long to unambiguously detect or rule out the presence
of old resolved (RGB) stars in this object. The data that we are
compiling on Cepheid stars in I~Zw~18 will be unique for probing the
properties of variable stars at metallicities that have never before
been probed.

\begin{acknowledgments}
Support for proposals \#9361 and \#10586 was provided by NASA through
a grant from STScI, which is operated by AURA, Inc., under NASA
contract NAS 5-26555.
\end{acknowledgments}

\end{document}